# *Logic Programming with Default, Weak and Strict Negations*

Susumu Yamasaki

*Department of Computer Science*
*Graduate School of Natural Science and Technology*
*Okayama University*
*Okayama 700-0082, Japan*
(*e-mail:* `yamasaki@momo.it.okayama-u.ac.jp`)

## Abstract

This paper treats logic programming with three kinds of negation: default, weak and strict negations. A 3-valued logic model theory is discussed for logic programs with three kinds of negation. The procedure is constructed for negations so that a soundness of the procedure is guaranteed in terms of 3-valued logic model theory.

## 1 Introduction

Negation in logic programming is so well studied that negation as failure is combined with SLD resolution from procedural views, and model theories are established in both 2-valued and 3-valued logic. There are approaches that add negations to logic programs, such as those to the answer set and strong negation (Gelfond and Lifschitz 1990; Pearce and Wagner 1991), but they are not really related to that proposed here. The present paper is relevant to 3-valued logic model theories (Fitting 1985; Kunen 1987; Shepherdson 1987; Przymusinski 1990; Van Gelder, Ross and Schlipf 1991; Baral and Subrahmanian 1993; Yamasaki and Kurose 2001). They are well constructed by means of monotonic mappings associated with programs involving negations. For default negation, some interpretations are made, with reference to the closed world assumption, the well-founded model and the 3-valued stable model (Ruiz and Minker 1997; Ruiz and Minker 1998). In contrast to multiple default negations, some usage(s) of different negation(s) with a single default negation is/are required.

For example, assume a scenario as follows. There are two places $a$ and $b$ as candidates for some event such that:

(i) If the place $a$ or $b$ is approved, then the event place is determined.
(ii) A conjunction of statements "$a$ is approved" and "$b$ is approved" is strictly negated.

  (iii) If a proposition $p$ holds, then $a$ is approved.
  (iv) If a statement "$a$ is approved" is strictly negated and an exclusion of a proposition "$a$ is preferred" is made, then $b$ is approved.
  (v) If $a$ is not preferred, then $b$ is preferred.

By means of strict negation (denoted by $\sim_s$), a strictly negated proposition is expressed. By means of weak negation (denoted by $\sim_w$), an exclusion of a proposition may be represented. Default negation (denoted by $not$) is used as is shown in the literature. (Note that the model-theoretic treatments of three negations are shown as below.) The first sentence (i) is translated into two clauses (a) and (b) of logic programming. The second sentence (ii) is represented by a goal (c). The third, fourth and fifth sentences (iii), (iv) and (v) are expressed by clauses (d), (e) and (f), respectively.

  (a) $determined \leftarrow approved(a)$
  (b) $determined \leftarrow approved(b)$
  (c) $\leftarrow approved(a), approved(b)$
  (d) $approved(a) \leftarrow p$
  (e) $approved(b) \leftarrow \sim_s approved(a), \sim_w preferred(a)$
  (f) $preferred(b) \leftarrow not\ preferred(a)$

In this paper, default ($not$), weak ($\sim_w$) and strict ($\sim_s$) negations are treated for logic programming, where "default" and "weak" negations are discussed in some of the literature, but the term "strictness" is adopted for the negation as below. In 3-valued logic, they map the truth values **t** ($true$), **u** ($undefined$) and **f** ($false$) to those values as follows.

$$not: \quad \mathbf{t} \mapsto \mathbf{f},\ \mathbf{u} \mapsto \mathbf{u},\ \mathbf{f} \mapsto \mathbf{t}.$$
$$\sim_w: \quad \mathbf{t} \mapsto \mathbf{f},\ \mathbf{u} \mapsto \mathbf{t},\ \mathbf{f} \mapsto \mathbf{t}.$$
$$\sim_s: \quad \mathbf{t} \mapsto \mathbf{f},\ \mathbf{u} \mapsto \mathbf{f},\ \mathbf{f} \mapsto \mathbf{t}.$$

Now the case of multiple occurrences of negations in parallel is examined. The default negation is already included in the established framework by means of negation as failure rule. The weak negation is interpreted as being "not true" (Schmitt 1986). Strict negation is of use to illustrate the constraint that a conjunction of propositions is contradictory.

We have a monotonic alternating mapping to denote the well-founded model for the program with default negation (Przymusinski 1990; Van Gelder, Ross and Schlipf 1991). By a simultaneous usage of default negation with weak or strict negation, some nonmonotonic mapping is associated with a logic program even in 3-valued logic. When weak and/or strict negations occur together with default negation in a logic program, the definition of a general fixpoint semantics is problematic. For model theory of the logic program involving multiple negations, procedural interpretations of negations may be effective:

  (i) Negation as failure for default negation
  (ii) Negation by weak failure (non-succeeding) for weak negation
  (iii) Negation by strict failure (failing) for strict negation

The primary goal of this paper is to present an abstract procedure containing rules for default, weak and strict negations. Its soundness is proven with respect to model theory, in the sense that the model is defined to be coherent with the procedure.

This paper is organized as follows. Section 2 is devoted to preliminaries for logic programs with three kinds of negation, as well as the alternating mapping. In Section 3, we provide a model theory and a procedure for the logic program with default, weak and strict negations. In this paper, for simplicity, we only deal with (ground) programs containing no variable. However, in Section 4 we make some remarks on the first-order logic programming and on hierarchical usages of negations.

## 2 Negation in Logic Programming

### 2.1 Logic Program and 3-Valued Interpretation

We deal with a finite (ground) set of clauses of the form $A \leftarrow L_1, \ldots, L_n$, where $A$ is an atom, and $L_1, \ldots, L_n$ are literals. The atom $A$ is referred to as the head of the clause, and the sequence $L_1, \ldots, L_n$ as its body. A goal is an expression of the form $\leftarrow L_1, \ldots, L_n$, where $L_1, \ldots, L_n$ are literals. When $n = 0$, the goal is denoted by $\square$. A literal is either $B$ (a positive literal), *not* $B$ (a negative literal with default negation), $\sim_w B$ (a negative literal with weak negation) or $\sim_s B$ (a negative literal with strict negation) for some atom $B$.

A definite program is a set of clauses containing no negative literal. A general logic program (LP, for short) is a set of clauses containing default negation, but neither weak negation, nor strict negation. The term of a "program" denotes a set of clauses with default, weak, and strict negations. In what follows, the set $B_P$ stands for the Herbrand base of a program $P$, which is the set of all atoms constructed by means of predicate and function symbols occurring in $P$.

*Definition 2.1*

If $I, J \subseteq B_P$ for a program $P$ such that $I \cap J = \emptyset$, then we say that the pair $(I, J)$ is a 3-valued Herbrand interpretation.

In the 3-valued Herbrand interpretation $(I, J)$, $I$ is regarded as a true set, and $J$ as a false set, while $B_P - (I \cup J)$ is regarded as the set of undefined atoms in $B_P$ for the truth value. The values **t**, **u** and **f** denote the truth values *true*, *undefined* and *false*, respectively, with respect to some interpretation. Assume a 3-valued Herbrand interpretation $(I, J)$. The evaluation $val_{(I,J)}(E)$ of the expression $E$ is defined with respect to $(I, J)$ as follows, where we assume an order among the truth values: **f** < **u** < **t**.

(1) Literal:

$$val_{(I,J)}(A) = \begin{cases} \mathbf{t} & (A \in I) \\ \mathbf{u} & (A \notin I \cup J) \\ \mathbf{f} & (A \in J) \end{cases} \qquad val_{(I,J)}(not\ A) = \begin{cases} \mathbf{f} & (A \in I) \\ \mathbf{u} & (A \notin I \cup J) \\ \mathbf{t} & (A \in J) \end{cases}$$

$$val_{(I,J)}(\sim_w A) = \begin{cases} \mathbf{f} & (A \in I) \\ \mathbf{t} & (A \notin I \cup J) \\ \mathbf{t} & (A \in J) \end{cases} \qquad val_{(I,J)}(\sim_s A) = \begin{cases} \mathbf{f} & (A \in I) \\ \mathbf{f} & (A \notin I \cup J) \\ \mathbf{t} & (A \in J) \end{cases}$$

(2) Body:

The value of the body $L_1, \ldots, L_n$, $val_{(I,J)}(L_1, \ldots, L_n)$, is the least one among the values of literals $L_1, \ldots,$ and $L_n$.

(3) Clause:

The clause is true if the value of the head is not less than that of the body, and false otherwise.

(4) Program:

The program $P$ is true if all the clauses have the truth value, and false otherwise.

We say that the interpretation $(I, J)$ is a 3-valued Herbrand model of the program $P$ if the value of the program is true, that is, $val_{(I,J)}(P) = \mathbf{t}$.

### 2.2 Alternating Mapping, Reviewed

Here we examine model theories in the 3-valued logic for programs. Before the examination, we review the model theory for LPs by means of alternating fixpoint semantics.

*Definition 2.2*
Let $K$ be a subset of the Herbrand base $B_P$ for an LP $P$. We define the set of clauses with respect to the set $K$ as follows.

$P[K] = \{A \leftarrow A_1, \ldots, A_m \mid$
$\quad \exists (A \leftarrow A_1, \ldots, A_m, not\ A_{m+1}, \ldots, not\ A_n) \in P : [A_{m+1}, \ldots, A_n \in K]\}.$

The mapping $S_P \colon 2^{B_P} \to 2^{B_P}$ is defined to be $S_P(K) = U_{P[K]} \uparrow \omega$, where we have the least fixpoint of the mapping $U_R$, $U_R \uparrow \omega = \cup_{i \in \omega} U_R \uparrow i$, for the definite program $R = P[K]$ as follows. The set $U_Q \uparrow i$ is defined for a definite program $Q$ by:

$$U_Q \uparrow i = \begin{cases} \emptyset & (i = 0), \\ U_Q(U_Q \uparrow (i-1)) & (i > 0). \end{cases}$$

The mapping $U_Q \colon 2^{B_Q} \to 2^{B_Q}$ is defined to be

$$U_Q(J) = \{B \mid \exists (B \leftarrow B_1, \ldots, B_n) \in Q : [B_1, \ldots, B_n \in J]\}$$

for a set $J \subseteq B_Q$.

Note that the mapping name "$U_Q$" is usually denoted by $T_Q$ (Lloyd 1993). However, the letter "$T$" is later reserved for the Herbrand interpretation $(T, F)$, so the mapping (name) $U_Q$ is adopted here. Because the mapping $U_Q$ is continuous, there is a least fixpoint $U_Q \uparrow \omega = \cup_{i \in \omega} U_Q \uparrow i$. $S_P$ is monotonic with respect to the subset inclusion ordering. We take the (alternating) mapping $\Theta_P : 2^{B_P} \to 2^{B_P}$ (Van Gelder 1993) to be $\Theta_P(K) = \overline{S_P(\overline{S_P(K)})}$, where the over-line stands for the operation taking the complement set with respect to the Herbrand base $B_P$. Since $S_P$ is monotonic, $\Theta_P$ is also monotonic. Therefore there is a fixpoint of $\Theta_P$, which is concerned with the 3-valued stable model. Note that its least fixpoint is directly in accordance with the well-founded model (Van Gelder, Ross and Schlipf 1991). Following the organization of (You and Yuan 1995), we give the definition of the 3-valued stable model (Przymusinski 1990) in terms of the alternating fixpoint.

*Definition 2.3*
Assume that $(S_P(J), J)$ is a 3-valued Herbrand interpretation. If $J$ is a fixpoint of $\Theta_P$, we say that $(S_P(J), J)$ is a 3-valued stable model of $P$.

## 3 Model Theory for Negations

In this section, we study a model theory for the program. The transformation method for LPs by means of *Definition* 2.2 is now extended to be applicable to the program.

*Definition 3.1*
Let $I, J, K$ be subsets of the Herbrand base $B_P$ for a program $P$. We define the set of clauses with respect to the sets $I, J$ and $K$ as follows.

$$P[I, J, K] = \{A \leftarrow A_1, \ldots, A_k \mid \\ \exists (A \leftarrow A_1, \ldots, A_k, not\ B_1, \ldots, not\ B_l, \sim_w C_1, \ldots, \sim_w C_m, \\ \sim_s D_1, \ldots, \sim_s D_n) \in P : \\ [B_1, \ldots, B_l \in I, C_1, \ldots, C_m \in J \text{ and } D_1, \ldots, D_n \in K]\}.$$

Note that the set of clauses $P[I, J, K]$ is a definite program for the mapping $U_{P[I,J,K]}$ to be applied, where we have seen the mapping $U_Q$ and the set $U_Q \uparrow \omega$ for the definite program $Q$ in the former section.

*Definition 3.2*
Assume a program $P$. A mapping $\Sigma_P : 2^{B_P} \times 2^{B_P} \times 2^{B_P} \to 2^{B_P}$ is defined to be $\Sigma_P(I, J, K) = U_{P[I,J,K]} \uparrow \omega$.

We see that if $I_1 \subseteq I_2$, $J_1 \subseteq J_2$ and $K_1 \subseteq K_2$, then $\Sigma_P(I_1, J_1, K_1) \subseteq \Sigma_P(I_2, J_2, K_2)$. In this sense, the mapping $\Sigma_P$ is monotonic.

*Definition 3.3*
Assume a program $P$. The semantic equations for $P$ are defined as follows.

$$\begin{cases} T &= \Sigma_P(F, \overline{T}, F), \\ F &= \Sigma_P(\overline{T}, \overline{T}, F), \end{cases}$$

such that $T \cap F = \emptyset$.

We refer to the equations of *Definition* 3.3 as the semantic equations for the program $P$. We do not always have any fixpoint of the semantic equations, because the transformation (caused by the semantic equations) is not monotonic, as is easily seen, for $T_0 \subseteq T_1$ and $F_0 = F_1$, $\Sigma_P(F_1, \overline{T_1}, F_1) \subseteq \Sigma_P(F_0, \overline{T_0}, F_0)$. For $T_0 = T_1$ and $F_0 \subseteq F_1$, $\Sigma_P(\overline{T_1}, \overline{T_1}, F_1) \subseteq \Sigma_P(\overline{T_0}, \overline{T_0}, F_0)$.

*Theorem 3.1*
Assume a program $P$. Also suppose for $P$ that

(a) $T \cap F = \emptyset$,
(b) $\Sigma_P(F, \overline{T}, F) = T$,
(c) $\Sigma_P(\overline{T}, \overline{T}, F) \subseteq \overline{F}$,
(d) $A \in F \Rightarrow$
    $\forall (A \leftarrow A_1, \ldots, A_k, \text{not } B_1, \ldots, \text{not } B_l, \sim_w C_1, \ldots, \sim_w C_m, \sim_s D_1, \ldots, \sim_s D_n) \in P$. [ some $A_i$ is in $F$, or [ some $B_{j_1}$ is in $T$, some $C_{j_2}$ is in $T$, or some $D_{j_3}$ is not in $F$ ] ]

for a pair $(T, F) \in 2^{B_P} \times 2^{B_P}$. Then the pair $(T, F)$ is a 3-valued Herbrand model of $P$.

*Proof*
Assume that $T \cap F = \emptyset$, that is, that the pair $(T, F)$ is a 3-valued Herbrand interpretation. Take any clause

$$A \leftarrow A_1, \ldots, A_k, \text{not } B_1, \ldots, \text{not } B_l, \sim_w C_1, \ldots, \sim_w C_m, \sim_s D_1, \ldots, \sim_s D_n$$

in $P$. We can examine the following cases.
(1) In the case that some $B_q$ is in $T$ for $1 \leq q \leq l$, some $C_r$ is in $T$ for $1 \leq r \leq m$, or some $D_s$ is in $\overline{F}$ for $1 \leq s \leq n$, the body of the clause is interpreted as **f** such that the clause has the value **t**.
(2) In the case that $B_1, \ldots, B_l$ are all in $F \subseteq \overline{T}$, $C_1, \ldots, C_m$ are all in $\overline{T}$, and $D_1, \ldots, D_n$ are all in $F$, we see by the definition of $\Sigma_P(F, \overline{T}, F) = U_{P[F, \overline{T}, F]} \uparrow \omega$ that:

$$[\ A_1, \ldots, A_k \in \Sigma_P(F, \overline{T}, F)\ ] \Rightarrow [\ A \in \Sigma_P(F, \overline{T}, F)\ ].$$

Because of (b), $\Sigma_P(F, \overline{T}, F) = T$. Hence the set $\Sigma_P(F, \overline{T}, F)$ denotes the set containing atoms evaluated as **t** in the interpretation $(T, F)$, and

$$[\ A_1, \ldots, A_k \text{ are } \mathbf{t} \Rightarrow A \text{ is } \mathbf{t}\ ].$$

It follows that the clause is evaluated as **t**.
(3) In the case that $B_1, \ldots, B_l$ are all in $\overline{T}$ such that some $B_q$ is not in $F$ for $1 \leq q \leq l$, $C_1, \ldots, C_m$ are all in $\overline{T}$, and $D_1, \ldots, D_n$ are all in $F$, we see by the definition of $\Sigma_P(\overline{T}, \overline{T}, F) = U_{P[\overline{T}, \overline{T}, F]} \uparrow \omega$ that:

$$[\ A_1, \ldots, A_k \in \Sigma_P(\overline{T}, \overline{T}, F)\ ] \Rightarrow [\ A \in \Sigma_P(\overline{T}, \overline{T}, F)\ ].$$

Because of (c), $\Sigma_P(\overline{T}, \overline{T}, F) \subseteq \overline{F}$ so that $\Sigma_P(\overline{T}, \overline{T}, F)$ denotes the set containing atoms evaluated as **t** or **u** in the interpretation $(T, F)$. If $A_1, \ldots, A_k$ are **t**, that is, in $T$, then $A$ is in $T$ as in case (2). As far as $A_1, \ldots, A_k \in \Sigma_P(\overline{T}, \overline{T}, F)$,

$$A_1, \ldots, A_k \text{ are } \mathbf{t} \text{ or } \mathbf{u},$$
$$\text{where at least one of them is } \mathbf{u} \;\; \Rightarrow \;\; A \text{ is } \mathbf{u}, \text{ that is, not } \mathbf{f}.$$

If $A_i \notin \Sigma_P(\overline{T}, \overline{T}, F)$ for some $A_i$ and $A_i \in F$, then the body of the clause is evaluated as **f**. If $A_i \notin \Sigma_P(\overline{T}, \overline{T}, F)$ for some $A_i$ and $A_i$ is in $\overline{F}$, then, in the case that the head $A$ is in $F$, the body of the clause is interpreted as **f**, because of the assumed condition (d). This concludes that the clause is interpreted as **t**. □

*Corollary 3.1*
If there is a fixpoint $(T, F)$ of the semantic equations for a program $P$ such that $T \cap F = \emptyset$, then it is a 3-valued Herbrand model of $P$.

*Proof*
Assume that the pair $(T, F)$ is a fixpoint of the semantic equations such that $T \cap F = \emptyset$. Then it satisfies the conditions (a), (b), (c) and (d) of *Theorem* 3.1. Thus it is a 3-valued Herbrand model of $P$. □

We have some illustrations for programs and fixpoints of their semantic equations.

*Example 3.1*
Assume a program $P = \{p \leftarrow \text{not } q, \sim_w r; r \leftarrow \sim_w p, \sim_s s\}$, where the expressions $p, q, r, s$ are atoms. There are fixpoints: $(\{p\}, \{q, r, s\})$, and $(\{r\}, \{p, q, s\})$.

*Example 3.2*
Assume a program $P = \{p \leftarrow \text{not } q, \sim_w r; r \leftarrow \text{not } r, \sim_s s\}$, where the expressions $p, q, r, s$ are atoms. There is a fixpoint $(\{p\}, \{q, s\})$.

*Example 3.3*
Assume a program $P = \{p \leftarrow \sim_w q; q \leftarrow p, \sim_s s\}$, where the expressions $p, q, s$ are atoms. There is no fixpoint.

## 4 Sound Proof Procedure

This section is devoted to an abstract procedure to define negations combined with SLD resolution, where its soundness is shown with respect to some model theory.

SLD resolution is a deduction formed by a rule to infer a goal

$$\leftarrow L_1, \ldots, L_{i-1}, M_1 \ldots, M_n, L_{i+1}, \ldots, L_m$$

from a goal $\leftarrow L_1, \ldots, L_{i-1}, A, L_{i+1}, \ldots, L_m$ and a clause $A \leftarrow M_1, \ldots, M_n$ in the given program. The default negation may be implemented by negation as failure combined with SLD resolution as follows (Kunen 1987; Shepherdson 1987; Shepherdson 1989; Lloyd 1993):

$$\leftarrow A \text{ succeeds} \;\; \Rightarrow \;\; \leftarrow \text{not } A \text{ fails},$$
$$\leftarrow A \text{ fails} \;\; \Rightarrow \;\; \leftarrow \text{not } A \text{ succeeds}.$$

The weak and strict negations are combined with SLD resolution and negation as failure:

(Negation by weak failure)

$$\leftarrow A \text{ succeeds} \Rightarrow \leftarrow \sim_w A \text{ fails},$$
$$\leftarrow A \text{ does not succeed} \Rightarrow \leftarrow \sim_w A \text{ succeeds}.$$

(Negation by strict failure)

$$\leftarrow A \text{ does not fail} \Rightarrow \leftarrow \sim_s A \text{ fails},$$
$$\leftarrow A \text{ fails} \Rightarrow \leftarrow \sim_s A \text{ succeeds}.$$

For the program $P$, we recursively define the relations $suc_P, fail_P \subseteq Goal$ for the set $Goal$ of goals, following the method (Kunen 1989). By the relations $suc_P(\leftarrow A)$ and $fail_P(\leftarrow B)$, we have the statements that the goal $\leftarrow A$ succeeds, and that the goal $\leftarrow B$ fails, respectively. The relations are to be the least set satisfying the following rule closure, where two rules (a) and (b) are contained for the rule closure to be compositional. A (possibly empty) sequence of literals is denoted by using letters like $G, G_1, G_2, \ldots$, where a sequence of literal sequences is also regarded as a sequence of literals. The rule is abstract, because it contains assumptions like "no relation $suc_P(\leftarrow A)$" and "no relation $fail_P(\leftarrow B)$". But it is referred to as a proof procedure, where it contains some operational views as SLDNF resolution. Note that the 3-valued Herbrand model is not always related to the least rule closure. For example, assume a program $P = \{p \leftarrow \sim_s q; q \leftarrow q\}$. The pair $(\{p\}, \{q\})$ is a model of $P$, while there is no relation $suc_P(\leftarrow p)$, nor $fail_P(\leftarrow q)$.

(a) $suc_P(\leftarrow G_1), suc_P(\leftarrow G_2) \Rightarrow suc_P(\leftarrow G_1, G_2)$.
(b) $fail_P(\leftarrow G) \Rightarrow fail_P(\leftarrow G_1, G, G_2)$.
(i) $suc_P(\Box)$.
(ii) $suc_P(\leftarrow G), (A \leftarrow G) \in P \Rightarrow suc_P(\leftarrow A)$.
(iii) $fail_P(\leftarrow A) \Rightarrow suc_P(\leftarrow not\ A)$.
(iv) no relation $suc_P(\leftarrow A) \Rightarrow suc_P(\leftarrow \sim_w A)$.
(v) $fail_P(\leftarrow A) \Rightarrow suc_P(\leftarrow \sim_s A)$.
(vi) no clause (in $P$) with $A$ in head $\Rightarrow fail_P(\leftarrow A)$.
(vii) for all clauses $A \leftarrow G$, $fail_P(\leftarrow G) \Rightarrow fail_P(\leftarrow A)$.
(viii) $suc_P(\leftarrow A) \Rightarrow fail_P(\leftarrow not\ A)$.
(ix) $suc_P(\leftarrow A) \Rightarrow fail_P(\leftarrow \sim_w A)$.
(x) no relation $fail_P(\leftarrow A) \Rightarrow fail_P(\leftarrow \sim_s A)$.

In the case that the program $P$ is finite, we implement another practical method as follows.

(1) Take a pair $(T, F)$ in $2^{B_P} \times 2^{B_P}$. On condition that $\mathbf{T} = \{suc_P(\leftarrow A) \mid A \in T\}$ and $\mathbf{F} = \{fail_P(\leftarrow B) \mid B \in F\}$, check whether the pair $(\mathbf{T}, \mathbf{F})$ is included in a rule closure.
(2) For all the pairs $(\mathbf{T}_1, \mathbf{F}_1), \ldots, (\mathbf{T}_n, \mathbf{F}_n)$ $(n \geq 0)$ like the pair $(\mathbf{T}, \mathbf{F})$ of (1), included in rule closures, respectively, we define a pair $(\cap_{1 \leq i \leq n} \mathbf{T}_i, \cap_{1 \leq i \leq n} \mathbf{F}_i)$ if $n \geq 1$, and get no pair if $n = 0$.
(3) The pair defined in step (2) is a required one, if it exits, and none otherwise.

The componentwise intersection of such pairs respectively included in rule closures is a required one, even if the program $P$ is countably infinite, because the componentwise intersection is contained in a rule closure.

Next we present soundness of the proof procedure with respect to model theory. For soundness of the proof procedure, we need a relation between the set $\Sigma_P(I, J, K)$ and the set derivable by SLD resolution.

*Definition 4.1*
Assume a program $P$. For the goal $\leftarrow A$ containing an atom $A$, the expression $Res_P(\leftarrow A)$ is the set of goals obtained by SLD resolution from the goal $\leftarrow A$, and is recursively defined as follows.

(i) The goal $\leftarrow A$ is in $Res_P(\leftarrow A)$.
(ii) If a goal $g$ is in $Res_P(\leftarrow A)$ and a goal $g'$ is obtained from the goal $g$ by SLD resolution, then the goal $g'$ is in $Res_P(\leftarrow A)$.

We define the set $n\text{-}Res_P(\leftarrow A)$ to be $\{\, g \in Res_P(\leftarrow A) \mid g \text{ contains only negative literals}\,\}$.

*Lemma 4.1*
Assume a program $P$. For an atom $A$,

$$A \in \Sigma_P(I, J, K) \Leftrightarrow \exists (\leftarrow not\ B_1, \ldots, not\ B_l, \sim_w C_1, \ldots, \sim_w C_m,$$
$$\sim_s D_1, \ldots, \sim_s D_n) \in n\text{-}Res_P(\leftarrow A):$$
$$[B_1, \ldots, B_l \in I, C_1, \ldots, C_m \in J, D_1, \ldots, D_n \in K].$$

*Proof*
We see that:

$$\begin{aligned}
A \in \Sigma_P(I, J, K) &\Leftrightarrow A \in U_{P[I,J,K]} \uparrow \omega \\
&\quad \text{(by the definition of } \Sigma_P) \\
&\Leftrightarrow \text{the goal } \square \text{ is obtained by SLD resolution from the goal } \leftarrow A \\
&\quad \text{with the definite program } P[I, J, K] \\
&\quad \text{(see (Lloyd 1993))} \\
&\Leftrightarrow \text{some goal} \\
&\quad \leftarrow not\ B_1, \ldots, not\ B_l, \sim_w C_1, \ldots, \sim_w C_m, \\
&\qquad \sim_s D_1, \ldots, \sim_s D_n \\
&\quad \text{is obtained by SLD resolution from the goal } \leftarrow A \\
&\quad \text{with the program } P \text{ such that } B_1, \ldots, B_l \in I,\ C_1, \ldots, C_m \in J, \\
&\quad D_1, \ldots, D_n \in K \\
&\quad \text{(by the relation of } P \text{ with } P[I, J, K]) \\
&\Leftrightarrow \exists (\leftarrow not\ B_1, \ldots, not\ B_l, \sim_w C_1, \ldots, \sim_w C_m, \\
&\qquad \sim_s D_1, \ldots, \sim_s D_n) \in n\text{-}Res_P(\leftarrow A): \\
&\quad [B_1, \ldots, B_l \in I, C_1, \ldots, C_m \in J, D_1, \ldots, D_n \in K] \\
&\quad \text{(by } Definition\ 4.1\text{).}
\end{aligned}$$

This concludes the proof. □

The following theorem states that if a goal $\leftarrow A$ succeeds, then there is a 3-valued Herbrand model $(T, F)$ such that $A \in T$. That is, the proof procedure is sound with respect to some associated model.

*Theorem 4.1*
Assume a program $P$ and a relation $suc_P(\leftarrow A)$. Then there is a 3-valued Herbrand model $(T, F)$ of $P$ such that:

(a) $A \in T$,
(b) $\Sigma_P(F, \overline{T}, F) = T$,
(c) $F \subseteq \overline{\Sigma_P(\overline{T}, \overline{T}, F)}$.

*Proof*
We here define the sets:
$$T = \{B \in B_P \mid suc_P(\leftarrow B)\},$$
$$F = \{C \in B_P \mid fail_P(\leftarrow C)\}.$$

We see as below that the pair $(T, F)$ is a 3-valued Herbrand interpretation of $P$:
Firstly a relation $R_P \subseteq B_P \times B_P$ is defined as follows:

$R_P(A_1, A_2) \Leftrightarrow$ both $suc_P(\leftarrow A_1)$ and $fail_P(\leftarrow A_1)$ hold, and
both $suc_P(\leftarrow A_2)$ and $fail_P(\leftarrow A_2)$ hold such that :
$suc_P(\leftarrow G)$ where $G$ contains *not* $A_2$.

We secondly see that: If we assume that $T \cap F \neq \emptyset$, that is, both the relations $suc_P(\leftarrow B)$ and $fail_P(\leftarrow B)$ hold for some atom $B$, then the relation $R_P$ is not empty. Assume that $B \in T \cap F$. Because the relation $suc_P(\leftarrow B)$ holds, we have a relation $suc_P(\leftarrow G_1, G_2, G_3)$ by applying the rule (ii) of the proof procedure backwards, for some goal $\leftarrow G_1, G_2, G_3$ in the set $n\text{-}Res_P(\leftarrow B)$, where:

$$\begin{aligned} G_1 &= \textit{not } B_1, \ldots, \textit{not } B_l, \\ G_2 &= \sim_w C_1, \ldots, \sim_w C_m, \text{ and} \\ G_3 &= \sim_s D_1, \ldots, \sim_s D_n. \end{aligned}$$

For the rules (iii), (iv) and (v) to be applied for the relation $suc_P(\leftarrow G_1, G_2, G_3)$,

$fail_P(\leftarrow B_1), \ldots, fail_P(\leftarrow B_l)$,
none of $suc_P(\leftarrow C_1), \ldots,$ none of $suc_P(\leftarrow C_m)$, and
$fail_P(\leftarrow D_1), \ldots, fail_P(\leftarrow D_n)$.

Since the relation $fail_P(\leftarrow B)$ holds then, it follows from backward applications of the rule (vii) of the proof procedure that we have a relation $fail_P(\leftarrow G_1, G_2, G_3)$ for the same goal $\leftarrow G_1, G_2, G_3$. By the rules (viii), (ix) and (x) to be applied for the relation $fail_P(\leftarrow G_1, G_2, G_3)$,

$suc_P(\leftarrow B_q)$ for some $B_q$,
$suc_P(\leftarrow C_r)$ for some $C_r$, or
none of $fail_P(\leftarrow D_s)$ for some $D_s$.

On the assumption that both the relations $suc_P(\leftarrow B)$ and $fail_P(\leftarrow B)$ hold, we see the relation $R_P(B, B_q)$ for the above $B_q$, by observing that

$suc_P(\leftarrow B_q)$ for some $B_q$ and $fail_P(\leftarrow B_q)$ such that:
$suc_P(\leftarrow G_1, G_2, G_3)$ where the sequence $G_1, G_2, G_3$ contains *not* $B_q$.

We thirdly take the notion of a "negative loop". We say that there is a negative loop for an atom $C$, if $R_P^+(C,C)$ for the transitive closure $R_P^+$ of the relation $R_P$. Because the transitive closure $R_P^+$ is not empty and the Herbrand base $B_P$ is finite, we have the relation $R_P^+(C,C)$ for some atom $C$. That is, there is a negative loop. If there is a negative loop for some atom $C$, then both the relations $suc(\leftarrow C)$ and $suc_P(\leftarrow G)$ hold for the sequence $G$ containing $not\ C$. But the relation $suc_P(\leftarrow G)$ cannot be recursively connected to the primitive relation $suc_P(\Box)$. By the least set condition of rule closure, this is a contradiction. That is, we are in contradiction to the assumption that $T \cap F \neq \emptyset$. Hence $T \cap F = \emptyset$.

(a) Assume a relation $suc_P(\leftarrow A)$. It follows from the construction of the set $T$ that $A \in T$.

(b) Assume that $B \in \Sigma_P(F, \overline{T}, F)$. By *Lemma* 4.1, there is some goal

$$\leftarrow not\ B_1, \ldots, not\ B_l, \sim_w C_1, \ldots, \sim_w C_m, \sim_s D_1, \ldots, \sim_s D_n \in \text{n-}Res_P(\leftarrow B)$$

such that
$$B_1, \ldots, B_l \in F,$$
$$C_1, \ldots, C_m \in \overline{T},\ \text{and}$$
$$D_1, \ldots, D_n \in F.$$

By the set definition of $F$, if $B_q \in F$ then $fail_P(\leftarrow B_q)$. As well, if $D_s \in F$ then $fail_P(\leftarrow D_s)$. By the set definition of $T$, if $C_r \in \overline{T}$ then there is no relation $suc_P(\leftarrow C_r)$. It follows from the rules (iii), (iv) and (v) of the proof procedure that

$$suc_P(\leftarrow not\ B_1, \ldots, not\ B_l, \sim_w C_1, \ldots, \sim_w C_m, \sim_s D_1, \ldots, \sim_s D_n).$$

Following finitely many applications of the rule (ii) of the proof procedure, $suc_P(\leftarrow B)$. That is, $B \in T$. Finally $\Sigma_P(F, \overline{T}, F) \subseteq T$. Conversely assume that $B \in T$. It follows from the definition that $suc_P(\leftarrow B)$. By finitely many backward applications of the rule (ii) of the proof procedure, we have a relation

$$suc_P(\leftarrow not\ B_1, \ldots, not\ B_l, \sim_w C_1, \ldots, \sim_w C_m, \sim_s D_1, \ldots, \sim_s D_n)$$

for a goal $\leftarrow not\ B_1, \ldots, not\ B_l, \sim_w C_1, \ldots, \sim_w C_m, \sim_s D_1, \ldots, \sim_s D_n \in \text{n-}Res_P(\leftarrow B)$ such that $fail_P(\leftarrow B_q)$ $(1 \leq q \leq l)$, none of $suc_P(\leftarrow C_r)$ $(1 \leq r \leq m)$ and $fail_P(\leftarrow D_s)$ $(1 \leq s \leq n)$. By the definitions of the sets $T$ and $F$, we see that

$$B_1, \ldots, B_l \in F,$$
$$C_1, \ldots, C_m \in \overline{T},$$
$$D_1, \ldots, D_n \in F.$$

By *Lemma* 4.1, $B \in \Sigma_P(F, \overline{T}, F)$. Therefore $T \subseteq \Sigma_P(F, \overline{T}, F)$. This completes the proof.

(c) Assume that $B \in \Sigma_P(\overline{T}, \overline{T}, F)$. By *Lemma* 4.1, there is some goal

$$\leftarrow not\ B_1, \ldots, not\ B_l, \sim_w C_1, \ldots, \sim_w C_m, \sim_s D_1, \ldots, \sim_s D_n \in \text{n-}Res_P(\leftarrow B)$$

such that
$$B_1, \ldots, B_l \in \overline{T},$$
$$C_1, \ldots, C_m \in \overline{T},\ \text{and}$$
$$D_1, \ldots, D_n \in F.$$

Table 1. Truth value for double negations

| $A$ | $not \sim_w A$ | $\sim_w not\ A$ | $not \sim_s A$ | $\sim_s not\ A$ | $\sim_w \sim_s A$ | $\sim_s \sim_w A$ |
|---|---|---|---|---|---|---|
| t | t | t | t | t | t | t |
| u | f | t | t | f | t | f |
| f | f | f | f | f | f | f |

By the set definition of $T$, if $B_q \in \overline{T}$ then there is no relation $suc_P(\leftarrow B_q)$. As well, if $C_r \in \overline{T}$ then there is no relation $suc_P(\leftarrow C_r)$. By the set definition of $F$, if $D_s \in F$ then $fail_P(\leftarrow D_s)$. Because we cannot apply any rule of (viii), (ix) and (x) of the proof procedure to the above goal, there is no relation $fail_P(\leftarrow not\ B_1, \ldots, not\ B_l, \sim_w C_1, \ldots, \sim_w C_m, \sim_s D_1, \ldots, \sim_s D_n)$. It follows that we cannot have the relation $fail_P(\leftarrow B)$. That is, $B \notin F$ ($B \in \overline{F}$). Therefore $\Sigma_P(\overline{T}, \overline{T}, F) \subseteq \overline{F}$. Finally $F \subseteq \Sigma_P(\overline{T}, \overline{T}, F)$.

(d) Assume any clause

$$A \leftarrow A_1, \ldots, A_k, not\ B_1, \ldots, not\ B_l, \sim_w C_1, \ldots, \sim_w C_m, \sim_s D_1, \ldots, \sim_s D_n$$

such that $A \in F$, that is, $fail_P(\leftarrow A)$. It follows from the rules (vi) and (vii), or the rules (viii), (ix) or (x) of the proof procedure, respectively that:

- $fail_P(\leftarrow A_i)$ for some $A_i$, that is, $A_i \in F$,
- $suc_P(\leftarrow B_{j_1})$ for some $B_{j_1}$, that is, $B_{j_1} \in T$,
- $suc_P(\leftarrow C_{j_2})$ for some $C_{j_2}$, that is, $C_{j_2} \in T$, or
- no relation $fail_P(\leftarrow D_{j_3})$ for some $D_{j_3}$, that is, $D_{j_3} \notin F$.

Hence the condition (d) of *Theorem* 3.4 is satisfied.
By the conditions (b), (c) and (d) with the consistency that $T \cap F = \emptyset$, it follows from *Theorem* 3.4 that the pair $(T, F)$ is a 3-valued Herbrand model of $P$. □

## 5 Concluding Remarks

Model theory and a sound proof procedure of the logic program with default, weak and strict negations are presented in the case of no occurrence of variables. Because the analysis of the semantic equations is not so easy, the procedural interpretations of the negations are studied. It is practical to lift it up to a first-order procedure by means of the safe rule for negations, that is, by means of the rule to allow only the ground negative literal containing no variable. The design of a procedure should be a problem for the possibly infinite Herbrand base. For a procedural definition with a non-safe rule, ideas on even the constructive negation (Stuckey 1991) may be needed, as well as on non-ground models considered for general logic programs (Gottlob, Marcus, Nerode, Salzer and Subrahmanian 1996; Yamasaki 1996).

On double usages of negation, Table 1 is obtained. The interpretations may involve the following aspects.

(i) $\sim_w not\ A$ and $\sim_w \sim_s A$ are "not false" (Schmitt 1986). They are complementary to the default negation $\sim_s A$ with reference to default negation, so that they are equivalent to $not\ \sim_s A$.

(ii) $\sim_s not\ A$ and $\sim_s \sim_w A$ are complementary to the weak negation $\sim_w A$ with reference to default negation, so that they are equivalent to $not\ \sim_w A$.

It is left to future studies, to analyze double negations and to define a procedure applied to such usages. In the paper (Przymusinski 1997), knowledge and belief operators are discussed in details, where their model theories are treated in logic programming with default negation. It might be a problem to examine whether the usages of weak and strict negations can be relevant to those operators.

## Acknowledgement

The author is indebted to the referees for revision of this paper.